\documentclass[12pt]{book}
\usepackage[dvips]{graphicx,color}
\usepackage{makeidx,tsukuba}
\usepackage{amssymb}
\makeauthorindex
\makeindex

\begin{document}
\BookTitle{\itshape The 28th International Cosmic Ray Conference}
\CopyRight{\copyright 2003 by Universal Academy Press, Inc.}
\pagenumbering{arabic}
\chapter{
An Estimate Of The Primary Mass Of Cosmic Rays At $10^{18}$ eV As Inferred From Volcano Ranch Data.}
\author{%
M. T. Dova$^1$, M. E. Mance\~nido$^1$, A. G. Mariazzi$^1$, T. P. McCauley$^2$ \& A. A. Watson$^3$\\
{\it
(1)Department of Physics,Universidad de La Plata,1900,Argentina\\
(2)Department of Physics,Northeastern University, Boston,MA 02115,USA\\
(3)Department of Physics and Astronomy,University of Leeds,Leeds LS2 9JT,UK}
}
\section*{Abstract} 
Accurate measurements of the lateral distribution of extensive air showers produced by cosmic rays of energy
greater than $10^{17}$ eV were made in the 1970s by Linsley.  At the time, the state of knowledge about the
best hadronic interaction model to use to describe such data prevented conclusions about the cosmic ray mass from being drawn.  
We have used a modern model, {\sc qgsjet98}, to infer the primary mass from these data, using the very careful records 
left by Linsley.  We find that at a median energy of $10^{18}$ eV , the data are well described by an iron-dominated 
composition (88 $\pm$ 6(stat) $\pm$ 18 (syst))\%.  
We discuss the systematic errors in this estimate that arise from model uncertainties and from the range of 
energies used in the work of Linsley. These data are used with the permission of the late John Linsley to whom this work is dedicated.
\section{Introduction}
Determining the mass composition above $10^{17}$ eV is an important and challenging measurement and has been tackled by several groups using different techniques.
These results have not all been consistent in their conclusions.
It is important to solve this problem because of its implications for cosmic ray models of origin, 
acceleration and propagation. 
We present a first interpretation of the precise Volcano Ranch measurements of the lateral distribution function using Monte Carlo tools that were not available when the data were recorded in the 1970s. 
The pioneering Volcano Ranch instrument was described by Linsley, in his various writings [2,3], 
with unusually detailed descriptions of his equipment, together with many shower details 
and a description of his data reduction methods.
\begin{figure}[t]
  \begin{center}
    \includegraphics[width=17.pc,height=17.pc]{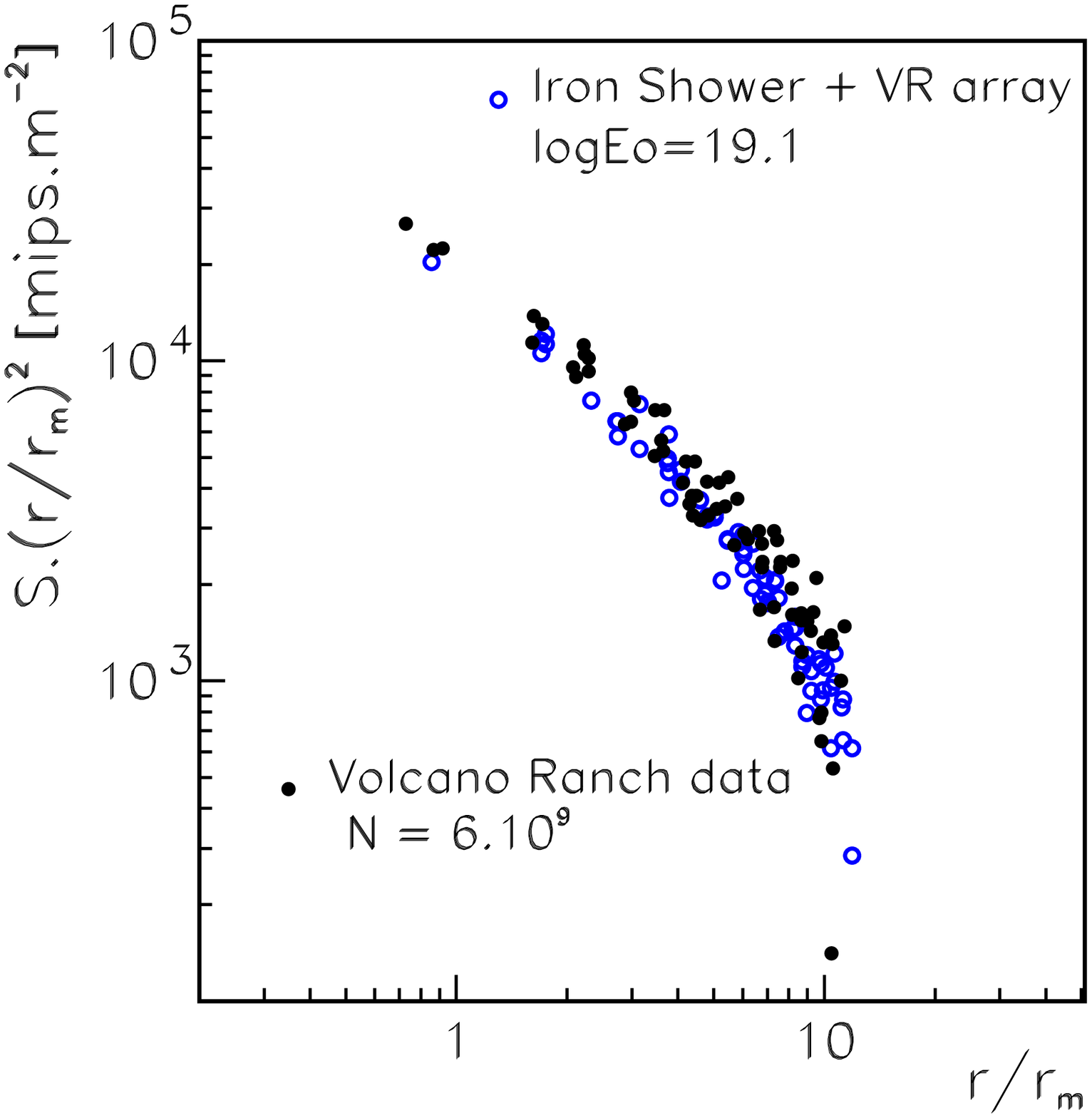}
    \includegraphics[width=17.pc,height=17.pc]{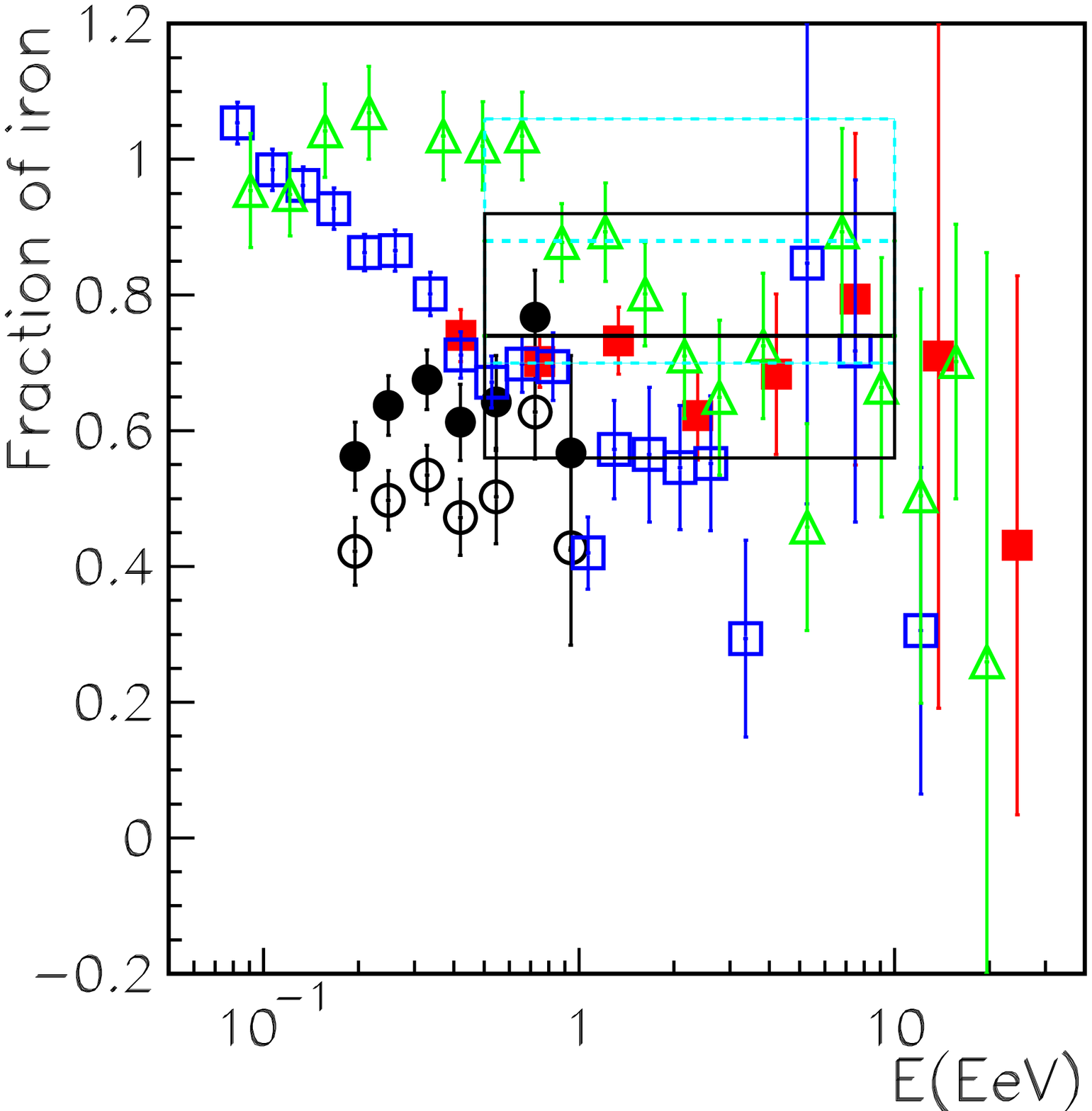} 
  \end{center}
\caption{
LEFT: Lateral distribution measurements compare with simulations.
\newline RIGHT: Fe fraction from various experiments: Fly's Eye ($\blacktriangle$), Agasa A100 ({\tiny $\blacksquare$}), Agasa A1 ({\tiny $\square$}) using a {\sc sibyll 1.5} [4] and references therein, Haverah Park [5] using {\sc qgsjet98} ({\large $\bullet$}) and  {\sc qgsjet01} ({\large $\circ$}). Mean composition determined in this paper with the corresponding error for the Volcano Ranch energy range using {\sc qgsjet98} (dashed rectangle) and an estimation of what it would result using {\sc qgsjet01} following [5] (solid line rectangle).
}
\end{figure}
\begin{figure}[t]
  \begin{center}
    \includegraphics[width=15.pc]{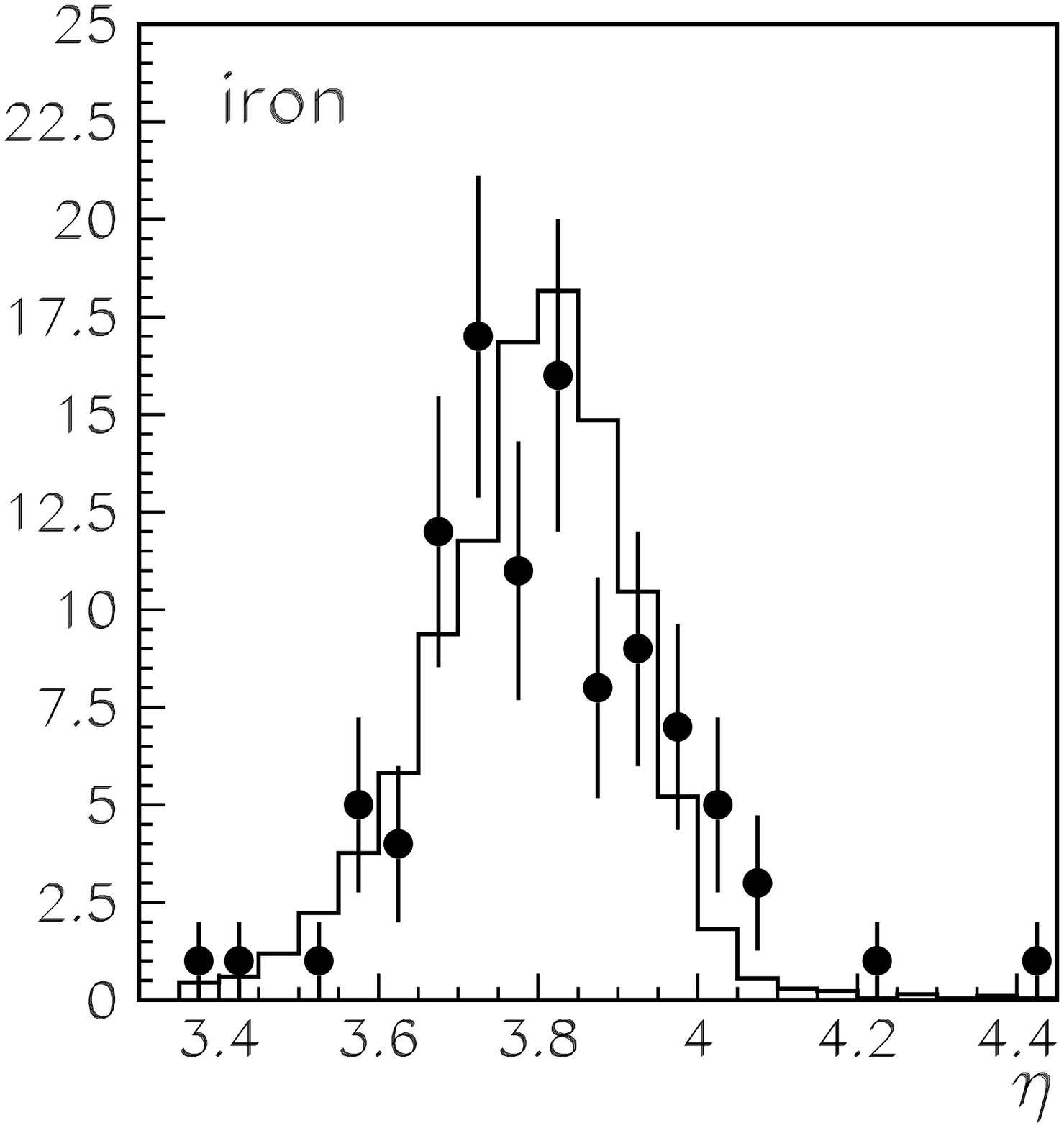}
    \includegraphics[width=15.pc]{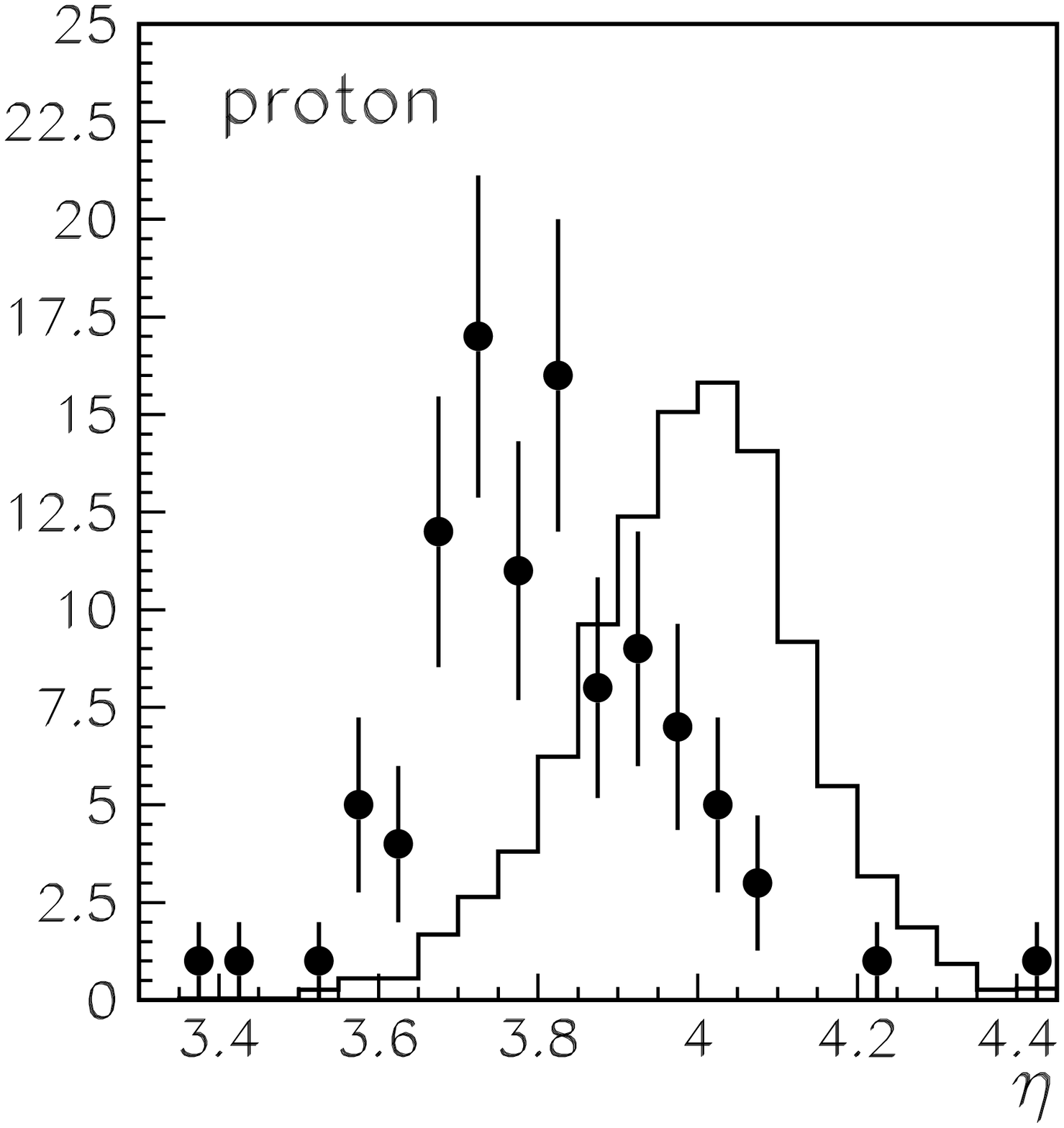} 
    \includegraphics[width=15.pc]{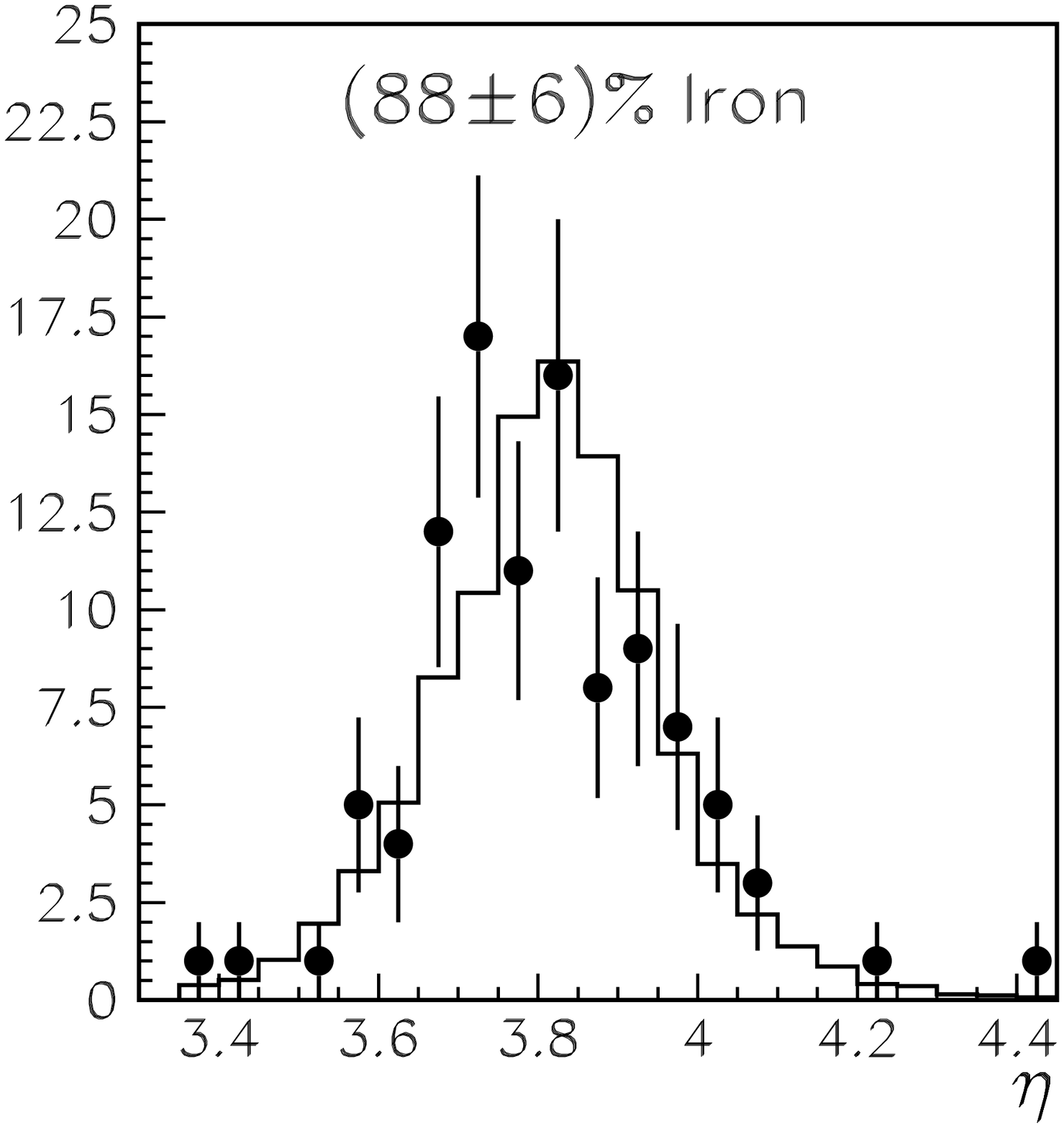} 
  \end{center}
\caption{The distributions of $\eta$ for pure iron (left), pure proton (right) and from the fit (bottom). Histograms: Monte Carlo; points: data.}
\end{figure}
\section{Derivation of the primary mass composition}
We have generated 1735 proton and iron showers with $\sec\theta=1.0-1.5$, and primary energy between 
$10^{17}$ eV and $10^{19}$ eV using the {\sc aires} code (version 2.4.0), with the hadronic interaction generator {\sc qgsjet98}. 
The results of the simulated showers were convolved with a simulation of the detector response made 
using {\sc geant4} to obtain the scintillator response [1].  
As a second step, each simulated shower was thrown up to 100 times on to a simulated Volcano Ranch array with 
random core positions in the range 0-150 m from the array center.
A comparison between lateral distribution measurements [2] and average $10^{19.1}$ eV iron showers 
including the scintillator response 
of the detectors in the Volcano Ranch array configuration is presented in Fig 1 (left). 
The agreement of the lateral distributions is good and gives confidence in the procedures used. 

A generalized version of the Nishimura-Kamata-Greisen (NKG) formula was used by Linsley to describe the lateral distribution [3]:
\begin{center}
\begin{equation}
S_{{\rm VR}}(r) = {\frac{N}{r_{m}^{2}}} \frac{\Gamma(\eta-\alpha)}{2\pi\Gamma(2-\alpha)\Gamma(\eta-2)} \left( {\frac{r}{r_{m}}} \right)^{-\alpha} \left( 1 + {\frac{r}{r_{m}}} \right)^{-(\eta-\alpha)} 
\nonumber
\end{equation}
\end{center}
From a subset of 366 showers detected with the array, the form of $\eta $ as a function of zenith angle $%
\theta $ and shower size $N$ was found to be [3] with $\alpha =1$: 
\begin{equation}
\langle \eta(\theta, N) \rangle = a + b(\sec \theta - 1) + c\:log_{10}({\frac{N}{10^{8}}})
\end{equation}
with $a = 3.88 \pm 0.054$, $b = - 0.64 \pm 0.07$, and $c = 0.07 \pm 0.03$.\\
We estimate the primary mass by a maximum likelihood fit of the best linear 
combination of pure iron and proton generated samples to match the data sample 
assuming a bi-modal composition. 
The probability of observing a particular number of events $d_i$ in a particular bin is given by 
$\exp^{-f_i}\,f_i^{d_i}/d_i!$ where $f_i$ is the predicted value for the number of events in this particular bin.\\
If we assume a bi-modal composition of proton and iron with proportions $P_{Fe}$ and $P_p$ then, $f_i=C(P_{Fe} + P_p)$ 
and $C$ is the overall normalization factor between numbers of data and Monte Carlo events.
The estimates of the proportions  $P_j$ are found by maximizing the total likelihood or its logarithm,
 $\ln(\pounds)=\Sigma d_i\ln(f_i)-\ln(d_i!)-f_i$.

In Fig 2 we show the Monte Carlo pure iron sample (left) and pure proton sample (right) with the corresponding Volcano Ranch data points.  
As can be seen from Fig 2 (right), the tail at large $\eta $, in the comparison with iron, indicates that a light component must be included to fit the experimental data.  
The best fit gives a mixture with (88 $\pm $ 6)\% of iron and a corresponding amount of protons. The resulting $\eta$ distribution is shown in Fig 2 (bottom) together with Volcano Ranch data.
The systematic error arising from our lack of knowledge of the energy distribution
of the events is estimated by repeating the fitting procedure with different energy spectra.
From these analyses we estimate a systematic error of 12\%.
The systematic error introduced by the choice of hadronic model is estimated to be 14\% [5].
\section{\protect\normalsize Discussion and Conclusions}
In Fig 1 (right) we present the fraction of iron for Fly's Eye, Agasa and Haverah Park using different hadronic interaction models. Also shown is the mean composition with the corresponding error we get for the Volcano Ranch energy range using {\sc qgsjet98} and an estimation of what we would get using {\sc qgsjet01} as it was done in Haverah Park [5]. 
Cosmic rays at Volcano Ranch are found to be compatible with mean fraction (88 $\pm$ 6(stat) $\pm$ 18 (syst))\% of iron 
in a bi-modal proton and iron mix, in the energy range $10^{17.7}$ eV to $10^{19}$ eV with mean energy $10^{18}$ eV.

The inconsistency between measurements of mass composition needs to be addressed
in order to make a more solid conclusion on the origin, acceleration or propagation of cosmic rays. 
This work attempts to help answer these important questions by re-interpreting the pioneering Volcano Ranch experiment, but in doing so has in some ways deepened the mystery of the origin of these particles.
Qualitatively there are significant differences between the estimates of composition at energies above $10^{17}$ eV.
So the questions appear far from being resolved. 
\section{References}
\vspace{\baselineskip}
\re
1.\ M. T. Dova et al, Proceedings of the XII ISVHECRI, CERN, 15-20 July 2002.
\re
2.\ J. Linsley, Proc. 13th ICRC, Denver, (1973) 3212.
\re
3.\ J. Linsley, Proc. 15th ICRC, Plovdiv,(1977) 56-62-89.
\re
4.\ B. R. Dawson et al, Astropart. Phys. 9 (1999) 331. 
\re
5.\ M. Ave et al., Astropart. Phys. 19 (2003) 61. 
\endofpaper
\end{document}